\documentclass{article}

\usepackage{arxiv}

\usepackage[utf8]{inputenc} 
\usepackage[T1]{fontenc}    
\usepackage{hyperref}       
\usepackage{url}            
\usepackage{booktabs}       
\usepackage{amsfonts}       
\usepackage{nicefrac}       
\usepackage{microtype}      
\usepackage{lipsum}		
\usepackage{graphicx}
\usepackage{doi}

\usepackage{multirow}
\usepackage[dvipsnames]{xcolor}
\usepackage{diagbox}

\title{A deep learning model for classification of diabetic retinopathy in eye fundus images based on retinal lesion detection}

\author{Melissa~delaPava \\
	Systems and Computer Engineering Department\\
	Universidad Nacional de Colombia\\
	Bogotá, Colombia, 111321. \\
	\texttt{medel@unal.edu} \\
	
	\And
	{Hernán Ríos} \\
	Fundación Oftalmológica Nacional\\
	Bogotá, Colombia, 110221 \\
	\texttt{hernanandresrios@gmail.com} \\
	
	\And
	{Francisco J. Rodríguez} \\
	Fundación Oftalmológica Nacional\\
	Bogotá, Colombia, 110221 \\
	\texttt{sfjrodriguez@fon.org.co} \\
	
	\And
	{Oscar J. Perdomo} \\
	School of Medicine and Health Sciences\\
	Universidad del Rosario\\
	Bogotá, Colombia, 111221\\
	\texttt{oscarj.perdomo@urosario.edu.co} \\
	
	\And
	{Fabio A. González} \\
	Systems and Computer Engineering Department\\
	Universidad Nacional de Colombia\\
	Bogotá, Colombia, 111321. \\
	\texttt{fagonzalezo@unal.edu.co} \\
	
}

\renewcommand{\shorttitle}{\textit{arXiv} Template}

\hypersetup{
pdftitle={A template for the arxiv style},
pdfsubject={q-bio.NC, q-bio.QM},
pdfauthor={David S.~Hippocampus, Elias D.~Striatum},
pdfkeywords={First keyword, Second keyword, More},
}

\begin{document}
\maketitle

\begin{abstract}
Diabetic retinopathy (DR) is the result of a complication of diabetes affecting the retina. It can cause blindness, if left undiagnosed and untreated. An ophthalmologist performs the diagnosis by screening each patient and analyzing the retinal lesions via ocular imaging. In practice, such analysis is time-consuming and cumbersome to perform. This paper presents a model for automatic DR classification on eye fundus images. The approach identifies the main ocular lesions related to DR and subsequently diagnoses the illness. The proposed method follows the same workflow as the clinicians, providing information that can be interpreted clinically to support the prediction. A subset of the \emph{kaggle EyePACS} and the \emph{Messidor-2} datasets, labeled with ocular lesions, is made publicly available. The \emph{kaggle EyePACS} subset is used as training set and the \emph{Messidor-2} as a test set for lesions and DR classification models. For DR diagnosis, our model has an area-under-the-curve, sensitivity, and specificity of $0.948$, $0.886$, and $0.875$, respectively, which competes with state-of-the-art approaches.
\end{abstract}


\section{Introduction}
\label{sec:intro}  
Diabetic retinopathy is an ocular disease that affects patients with undiagnosed, untreated, or undertreated diabetes mellitus. DR causes damage to the vessels inducing leakage of fluid within the retina, exudates and intraretinal hemorrhages. If the disease progresses it may lead to decreased vision and even blindness~\cite{Sengar2017}. Almost 40\% of diabetic patients have DR, from which about 5\% of patients face vision-threatening complications \cite{Antal2014}.

An ophthalmologist performs the DR diagnosis through a meticulously visual inspection of the retina according to the ICDR grading system \cite{Wilkinson2003}, which determines the DR level according to the presence of retinal lesions such as microaneurysms (MA), hemorrhages (H), exudates (EX), cotton wool spots (CWS), intraretinal microvascular abnormalities (IRMA), venous beading (VB), and neovascularization (NV) \cite{Chudzik2018, Sengar2017}. The initial clinical signs of DR are microaneurysms which allow leakage of blood from the affected capillaries, they appear as small reddish dots on the superficial retinal layers. Microaneurysms tend to have weak walls that can break, leading to hemorrhages of variable size and shape \cite{Paing2016}. Damaged retinal vessels allow lipoproteins precipitates to leak, which can be seen as yellowish exudates of irregular shape. These exudates exhibit a characteristic brightness in comparison to the contrast presented in microaneurysms and hemorrhages. CWS are lesions caused by retinal nerve fiber layer ischemia and emerge as yellowish-white spots with irregular edges. Signs of more advanced DR include irregular constriction and dilation of venous vessels in the retina, known as venous beading \cite{Chen2018a}, as well as neovascularization, which appears as red fronds of abnormal vascular networks. These vessels may show up arranged in a radiating pattern or without a distinct pattern \cite{talks2015}.

As much as 95\% of the cases of vision loss and blindness can be prevented with regular screening and appropriate management \cite{Bhaskaranand2019}. Despite the importance of a timely diagnosis fewer than a half of the patients with DR are aware of their condition \cite{Sengar2017}. Due to the asymptomatic nature of the disease in its early stages \cite{Mookiah2013}, patients should be screened with a fundus exam, to look for early signs of the disease and make a timely detection of the illness \cite{Beagley2014, Paing2016}. However, access to specialized care by an ophthalmologist is limited for some populations, given that most of the ophthalmologists in the world are concentrated mainly in urban areas and big cities. The development of an automated detection system for DR could improve access to specialized care by reducing the time, cost, and effort of screening \cite{Resnikoff2019, Orlando2018, Toledo2020}. In addition, the diabetic population is expected to increase 54\% by 2030, while the projected increase of ophthalmologists is only 2\%. Thus, the need to integrate methods to automate the screening process responds to these challenges in the present and future diabetic retinopathy panorama \cite{VanderHeijden2018}.

Dealing with the previous issues, we propose a model for the automatic detection of DR that initially relies on the identification of its ocular related lesions to later diagnose this illness. The proposed method is different from state-of-the-art models based on Convolutional Neural Networks (CNN), because it follows the clinician workflow, this approach has the advantage of providing additional information regarding the lesions found in the input image that give an improved interpretability. The experimental results also showed that this strategy also improves the overall classification performance of the system.

The main three contributions of this paper are as follows: a fine-grained lesions annotations for microaneurysms, hemorrhages, cotton wool spots, venous beading, neovascularization and exudates; global labels for ocular diseases and referable conditions for $3209$ images from the kaggle EyePACS dataset and the complete Messidor-2 dataset; and a strategy for the detection of DR that relies on the identification of its related retinal abnormalities.

The paper is organized as follows: Section~\ref{sec:previous_work} presents a review of the state-of-the-art including the most representative approaches that classify DR based on retinal lesions identification. Section~\ref{sec:databases} presents a description of the features of our new fine-grained labels for lesions of the retina related to DR and a comparison with the available databases. The proposed method is described in detail in Section~\ref{sec:method}, the results are summarized and discussed in Section~\ref{sec:results}, and conclusions and future work are presented in Section~\ref{sec:conclusion}.

\section{Previous work}
\label{sec:previous_work}
In this section, we present an overview of the most representative methods for the identification of DR based on the detection of individual and multiple ocular lesions.
The most common approach contains two stages: a first step extracts manually a set of features from DR-related lesions, and then a model uses them to classify or grade the DR. 

Sharif et al. presented a model that combines independent component analysis with a curve fitting technique to remove retinal blood vessels and the optic disc from eye fundus images. A feature set of DR lesions is build, which includes the count number of EX, H, MA, mean, and standard deviations of candidate regions on the images. Then, a multi-class Gaussian Bayes classifier and a multi-SVM are trained with the feature set to grade DR~\cite{Sharif2019}. Similarly, Abdelmaksoud et al. reported a known segmentation method named as U-Net to discriminate among EX, MA, H, and blood vessels. Then a set of features as co-occurrence matrix, areas, and bifurcation points count are calculated to grade DR using a SVM~\cite{Abdelmaksoud2021}. Paing et al. proposed a method that uses histogram matching, morphological opening, and canny method to segment blood vessels, EX and MA. Some features as area and counts are estimated to classify the stages of DR using a customized artificial neural network~\cite{Paing2016}. Akram et al. presented a method that initially removes blood vessels and optic disc. Then, a set of morphological descriptors such as shape, color, and statistical features are used to train a weighted combination of multivariate m-Mediods and a Gaussian Mixture Model to identify potential candidates for MA, H, and EX. Finally, the fundus image is graded on a DR scale according to medical conditions that include the type and count of lesions~\cite{UsmanAkram2014}.

Alternatively, some researchers have proposed models to automatically perform feature extraction from DR-related lesions and use them to classify the illness. Yang et al. presented a two-stages CNN-based algorithm trained with image patches and a weighted lesion map on the input, which can detect DR severity and its related lesions~\cite{Ghafoorian2017b}. Wang et al. designed a hierarchical multi-task deep learning framework for the classification of DR severity and DR-related lesions such as MA, H, CWS, VB, NV, and others~\cite{Wang2020a}. Similarly, Zago et al. explored a lesion localization model using two CNN patch-based approaches. Additional information such as a lesion probability map and the maximum value of the probability map are used to classify DR~\cite{Zago2020}. Recently, Zhou et al. partially released fine-grained annotations of pixel-level and image-level lesions related to DR from the FGADR dataset. They reported individual modules for the segmentation of lesions to extract features and to automatically grade DR~\cite{Zhou2021}.

Our model does not require segmentation of retinal structures as a preprocessing step or any other special preprocessing of the images, unlike most of the methods in literature for the detection of DR based on lesions. Besides, we did not reannoted the DR labels, our DR detection model is evaluated with the already publicly available so it can be compared with state-of-the-art methods and all the additional labels of DR lesions used to train the model are going to be publicly available.

\section{Dataset description}
\label{sec:databases}
The kaggle EyePACS is a free real-world set of high-resolution photos of the posterior pole of the retina. The kaggle EyePACS dataset comprises $88702$ macula-centered eye fundus images acquired with different types of cameras and under a variety of imaging conditions. These images were labeled using a scale of 0, 1, 2, 3, 4, which stand for no DR, mild, moderate, severe and proliferative DR respectively~\cite{kaggle}.

The Messidor-2 dataset is a public dataset that contains 1748 macula-centered eye fundus images that were acquired with a 45-degree field of view and the sizes are ranged between $1440\times960$ and $2304\times1536$ pixels. Abramoff et al. released the images with a binary classification for presence or absence of DR~\cite{messidor_labels}.

These public datasets were used to create two new individual datasets with fine-grained labels of six ocular lesions used for ophthalmologists to diagnose DR. The construction details are explained as follows: 

\emph{Dataset construction:} an ophthalmologist with supra-specialty in the retina from the Fundacion Oftalmologica Nacional of Colombia selected $3209$ retinal images from the kaggle-EyePACS dataset. The ophthalmologist performed the manual choice of these images based on the quality of fundus images and the presence of DR-related lesions. The Messidor-2 dataset was also analysed by the specialist who determined that $1689$ images are suitable for lesion level annotation. The expert labeled the images with six ocular findings: aneurysms, hemorrhages, cotton wool spots, venous beating, neovascularization, and exudates. In addition, these images were manually annotated with: binary labels for referable and non-referable patients and the grade of diabetic macular edema (DME) in a scale 0 to 3 for no DME, mild, moderate, and severe.

\emph{Annotation criteria:} the whole images were evaluated in their entire area as follows: first, a meticulous analysis looking for any lesions on the optic nerve and the macular region is performed. Then, the extra-macular retina and outside the vascular arches areas are examined in detail to find possible DR-related lesions. Finally, each image was classified in a fine-grained way as positive for that specific finding if at least one ocular lesion was identified. Regardless of whether it was a single injury or many, or whether it affected the center of the macula or its periphery.  In addition, the images were labeled as DME in an image-level way according to the ETDRS scale. Also, images were classified as a referable image, if at least an ocular lesion was present in all image areas, regardless of which one.

\begin{table}[!hb]
\caption{Distribution of DR-related ocular lesions and image-level labels from the customized Kaggle EyePACS (training test) and Messidor-2 (test set) data sets.}
\begin{tabular}{c|c|c|c|c|c|c||c|c|c|}
\cline{2-10}
& \multicolumn{6}{c||}{Ocular lesions labels} & \multicolumn{3}{c|}{Image-level labels}\\ 
\cline{1-10} 
\multicolumn{1}{|c|}{Set of images} & \multicolumn{1}{c|}{MA} & \multicolumn{1}{c|}{H} & \multicolumn{1}{c|}{CWS} & \multicolumn{1}{c|}{VB} & \multicolumn{1}{c|}{NV} & \multicolumn{1}{c||}{EX} & \multicolumn{1}{c|}{DR} & \multicolumn{1}{c|}{REF}  & \multicolumn{1}{c|}{DME}\\ \hline
\multicolumn{1}{|l|}{\multirow{2}{*}{Training set}} & 1719 & 1418 & 653 & 720 & 140 &  1812 & 2016 & 2550 & 1812\\ 
\multicolumn{1}{|l|}{} & 53.55\% & 44.17\% & 20.34\% & 22.42\% & 4.36\% & 56.44\% & 62.80\% & 79.44\% & 56.44\%\\ 
\hline
\multicolumn{1}{|l|}{\multirow{2}{*}{Test set}} & 876 & 443 & 133 & 47 & 15 &  196 & 377 & 893 & 196\\ 
\multicolumn{1}{|l|}{} & 51.86\% & 26.22\% & 7.87\% & 2.78\% & 0.89\% & 11.60\% & 22.32\% & 52.87\% & 11.60\%\\ 
\hline
\end{tabular}
\label{tab:dataset_distribution}
\end{table}

A summary with the number of images and percentage per finding and image-level for the customized kaggle EyePACS dataset which is used as training set and for the Messidor-2 dataset, which is used as test set is presented in Table \ref{tab:dataset_distribution}. 

The DR-related findings of MA, H, and CWS are the most common lesions as shown in Table~\ref{tab:dataset_distribution}. Moreover, these ocular findings are related to appear in the initial stages of subjects with DR. On the other hand, VB and NV findings presented few examples during the screening, but also these findings are scarce and rare to find because they are common in an advanced grade of DR. Finally, the Referable condition was the most common image-level labels among the datasets.

The comparison of our dataset with others publicly available with fined grained labels of DR related lesion is presented in Table \ref{tab:data_bases}. It can be noticed that the available datasets have fewer examples of DR related lesions, this is due to labeling images at a lesion level is costly, tedious and time consuming \cite{Orlando2018}. We hope that this new set of labels open opportunities to improve and develop new approaches for the detection of retina lesions caused by DR and for the detection of DR based on them, which leads to more relatable models for the clinicians.
\begin{table}[!ht]
\centering
\caption{Comparison of DR-related lesion labels from public databases. The DR-related lesion are: microaneurysm (MA),  hemorrhages (H), cotton wool spots (CWS), venous beading (VB), neovascularization (NV), exudates (EX), and other lesions.}
\begin{tabular}{|p{4.7cm}|>{\centering\arraybackslash}p{1cm}|>{\centering\arraybackslash}p{1cm}|>{\centering\arraybackslash}p{1cm}|>{\centering\arraybackslash}p{1cm}|>{\centering\arraybackslash}p{1cm}|>{\centering\arraybackslash}p{1cm}|>{\centering\arraybackslash}p{1cm}|>{\centering\arraybackslash}p{1cm}|>{\centering\arraybackslash}p{1cm}|}
\hline
\diagbox{Database}{Ocular lesions} & MA & H  & CWS & VB & NV & EX & Others\\
\hline
DRIVE         &   &  X &    &    &    & X   & X \\
DIARETDB0     & X  & X & X  &    &  X & X   &   \\
DIARETDB1     & X  & X & X  &    &    & X   &   \\
STARE         & X &    &    &    &  X &     & X \\
IDRiD         & X & X  & X  &    &    & X   &   \\
FGADR         & X & X  & X  &    & X  & X   & X \\
e-ophtha      & X &    &    &    &    & X   &   \\
Our dataset   & X & X  & X  &  X & X  & X   &   \\
\hline
\end{tabular}
\label{tab:data_bases}
\end{table}

\section{Method}\label{sec:method}
\begin{figure}[!hb]
    \centering
    \includegraphics[scale=0.42]{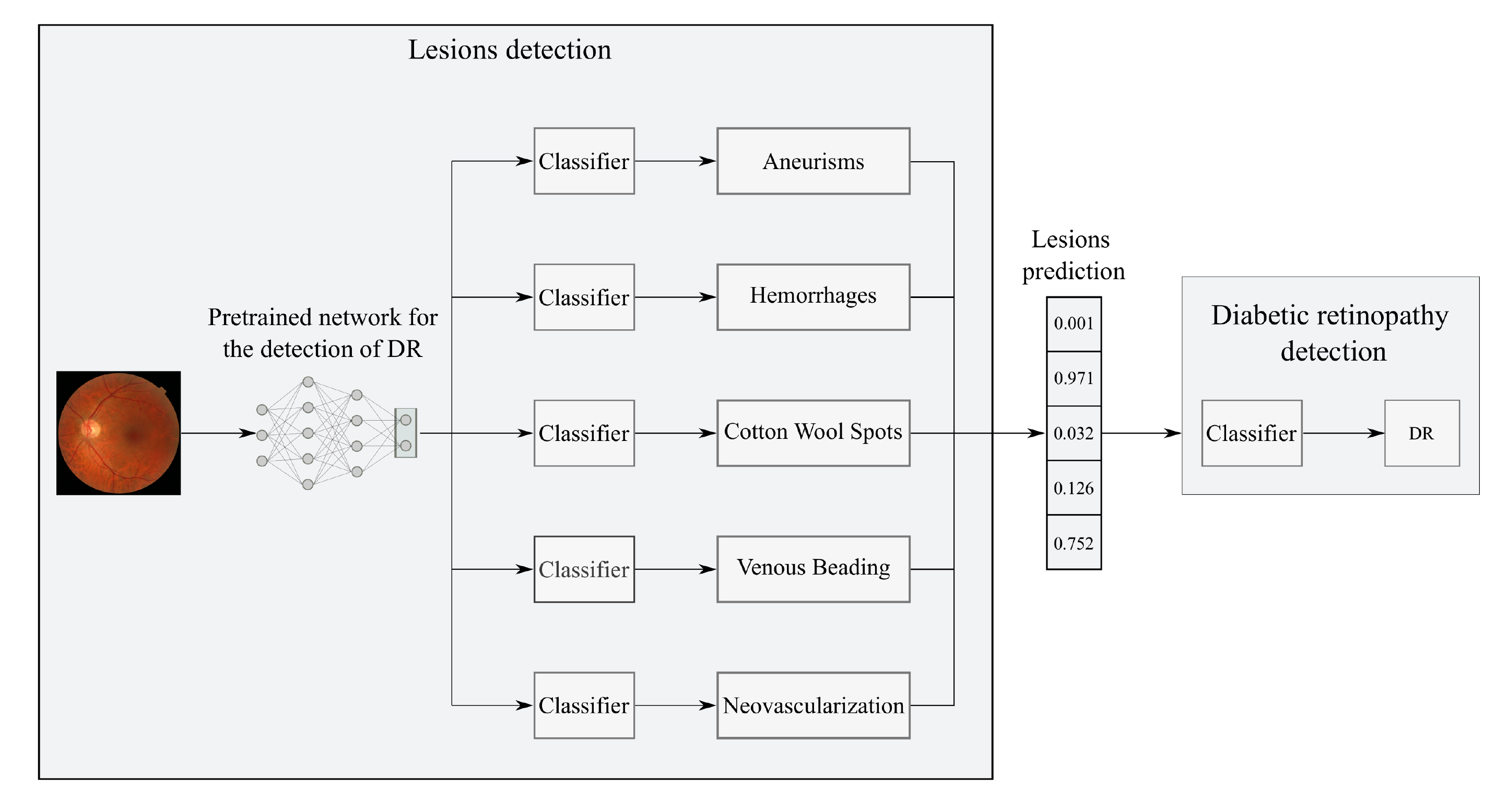}
    \caption{An overview of the proposed method for ocular lesions detection and DR classification. The model is organized in three consecutive stages as follows: lesion detection (first block), lesion prediction (second block) and DR detection (third block). }
    \label{fig:model}
\end{figure}

The pipeline of our proposed method is depicted in Figure~\ref{fig:model}. The method has two main stages, lesion detection and DR classification. 

Initially, the eye fundus images are preprocessed to resize them to $512 \times 512$ pixels and remove the black edges. The lesion detection stage is based on a transfer learning strategy using the weights of pre-trained models for DR detection reported by He et al.~\cite{He2021}. The base models correspond to the embedding of two attention blocks into different backbones networks, namely, DenseNet121, Xception,  ResNet50, and MobileNet for DR grading. Multiple classifiers are trained using the features extracted from the base networks and the fine-grained labels of the kaggle EyePACS dataset for the detection of DR-related lesion. The best overall performing method is selected as the classifier for the detection of lesions.

In the second stage the predictions of the lesions are concatenated to train multiple classifiers for the DR classification task. The best results are reported and compared with state-of-the-art methods using the Messidor-2 dataset as test set. The models and datasets used in this work will be released in a public repository to ensure reproducibility and encourage other researchers to create new methods.

The classifiers and the parameters explored in both stages are:
\begin{itemize}
    \item Support vector machine (SVM): the parameters explored are the regularization parameter (C), different kernels and the gamma $(\gamma)$ kernel coefficient.
    \item Gaussian Process (GP): the kernels `RBF' and `Mattern' are evaluated. The kernel parameters explored are lower length scale bound, upper length scale bound, lower noise level and upper noise level.
    \item Multilayer Perceptron (MLP): the number of layers, number of neurons, activations and learning rate are explored.
\end{itemize}

\section{Results and Discussion}\label{sec:results}
\subsection{Lesions automatic detection}
A systematic exploration and evaluation of the detection of DR-related lesions using the different classifiers and the features obtained from the pre-trained models was performed. After an initial exploration, the best overall results are obtained using as backbone the {densenet121\_CAB\_EyePACS} model as feature extractor. The results of the detection of MA, H, CWS, VB, and NV using this model are reported in Table~\ref{tab:findings_results}, where the best performing classifiers are the GP regressor and the MLP. The MA and VB findings presented greater difficulty in comparison to H and NV findings to be detected. This may be because, hemorrhages are bigger and neovascularizations are usually accompanied by other lesions what can help the model to detect them since it appears in the latter stages of the disease. 
\begin{table}[!ht]
\centering
\caption{Results of the detection of DR-related ocular lesions using the densenet121\_CAB\_EyePACS backbone model as a feature extractor.}
\begin{tabular}{|c|p{1.2cm}cc>{\centering\arraybackslash}p{1.5cm}>{\centering\arraybackslash}p{1.5cm}|}
\hline
Finding              & Metric & GP classifier &  GP regressor & SVM    & MLP    \\ \hline
\multirow{3}{*}{MA} & AUC & 0.7377 & 0.7865 & 0.7516 & 0.8055 \\
& Sp     & 0.8794        & 0.7023        & 0.8733 & 0.7072 \\
& Se     & 0.5959        & 0.7141        & 0.5833 & 0.7237 \\\hline
\multirow{3}{*}{H}   & AUC    & 0.8457        & 0.9332        & 0.8503 & 0.9317 \\
                     & Sp     & 0.9622        & 0.8472        & 0.9775 & 0.8370 \\
                     & Se     & 0.7291        & 0.8600        & 0.7088 & 0.8826 \\ \hline
\multirow{3}{*}{CWS} & AUC    & 0.8791        & 0.8825        & 0.6884 & 0.9021 \\
                     & Sp     & 0.7596        & 0.7622        & 0.9485 & 0.8303 \\
                     & Se     & 0.8496        & 0.8571        & 0.4135 & 0.8496 \\ \hline
\multirow{3}{*}{VB}  & AUC    & 0.7525        & 0.7299        & 0.5348 & 0.7983 \\
                     & Sp     & 0.6461        & 0.6565        & 0.9098 & 0.8026 \\
                     & Se     & 0.6808        & 0.6808        & 0.1702 & 0.5957 \\ 
                     \hline
\multirow{3}{*}{NV}  & AUC    & 0.9401        & 0.9693        & 0.6205 & 0.9685 \\
                     & Sp     & 0.8393        & 0.9074        & 0.9164 & 0.9127 \\
                     & Se     & 0.8666        & 0.9333        & 0.3333 & 0.9333 \\ \hline
\end{tabular}
\label{tab:findings_results}
\end{table}

\subsection{DR detection using lesions prediction}
The proposed method trained with a reduced set of training samples shows competitive performance in AUC and consistent results in sensitivity and specificity evaluated in Messidor-2 when compared to the baseline models trained with larger datasets for the DR detection task as shown in Table~\ref{tab:dr_results_comparison}. Despite some state-of-the-art methods have better performances, they only classify DR, unlike our approach that also provides information about the DR-related lesions, which provides interpretability and makes the DR detection process more familiar for clinicians, since it follows a workflow similar to their own.
\begin{table}[!htbp]
\centering
\caption{Comparison of proposed method with state-of-the-art DR detection methods tested on Messidor-2 dataset. The top 3 results for each metric are marked in bold$^{(1)}$, italic$^{(2)}$, and underline$^{(3)}$ respectively.}
\begin{tabular}{|>{\centering\arraybackslash}p{2.5cm}|p{4.5cm}|>{\centering\arraybackslash}p{1.55cm}|>{\centering\arraybackslash}p{1.55cm}|>{\centering\arraybackslash}p{1.65cm}|}
\hline
\shortstack{Training set  \\ (set of images) }  & Method & AUC & Sensitivity & Specificity\\
\hline
EyePACS train (35126) & densenet121\_CAB\_DDR~\cite{He2021}    & 0.765 & 0.359 & \textbf{0.990$^{(1)}$}\\
EyePACS train (35126)& densenet121\_CAB\_EyePACS~\cite{He2021} & 0.782 & 0.395 & \textit{0.954$^{(2)}$}\\
EyePACS train (35126)& resnet50\_CAB\_EyePACS~\cite{He2021}    & 0.884 & 0.753 & 0.801\\
EyePACS train (35126)& xception\_CAB\_DDR~\cite{He2021}        & 0.908 & 0.877 & 0.861\\
EyePACS train (35126)& xception\_CAB\_EyePACS~\cite{He2021}    & 0.908 & \underline{0.880}$^{(3)}$ & 0.855\\
EyePACS custom (57146) & Voets et al.~\cite{Voets2018}         & 0.800 & 0.737 & 0.697\\
EyePACS custom (28102) & Zhou et al.~\cite{Zhou2018}           & \textbf{0.960$^{(1)}$} & -     & - \\
EyePACS custom (75137) & Gargeya and Leng \cite{Gargeya2017}   & \underline{0.940}$^{(3)}$ & \textbf{0.930$^{(1)}$} & 0.870\\
EyePACS custom (3209) & our model with MLP & \textit{0.948$^{(2)}$} & \textit{0.886$^{(2)}$} & \underline{0.875}$^{(3)}$\\
\hline
\end{tabular}
\label{tab:dr_results_comparison}
\end{table}

\section{Conclusion} \label{sec:conclusion}
We presented a strategy to improve the performance of a pre-trained model for the detection of DR by using it as feature extractor to classify DR related lesions, and use them to identify the illness. The best results are obtained using a MLP as classifier, which leads to an end-to-end model that is easier to handle and can be further improved. 
The most remarkable aspect of our proposed method is the combination of domain knowledge from ophthalmologists and the deep learning versatility to support medical decision making. Moreover, the development of models that includes DR-related lesions and the clinical workflow to perform the ocular disease diagnosis may be a good solution to gain greater acceptance and possible use in real-world applications by clinical staff.
Finally, the validation of our method with a large number of images and other ocular diseases will be studied in a future work.

\bibliography{template} 
\bibliographystyle{ieeetr} 






\end{document}